# Implementing Quantum Generative Adversarial Network (qGAN) and QCBM in Finance

Santanu Ganguly
*Quantum Networks*
Photonic Inc.
London, United Kingdom
ORCID: 0000-0003-0141-9228

*Abstract*—Quantum machine learning (QML) is a cross-disciplinary subject made up of two of the most exciting research areas: quantum computing and classical machine learning (ML), with ML and artificial intelligence (AI) being projected as the first fields that will be impacted by the rise of quantum machines. Quantum computers are being used today in drug discovery, material & molecular modelling and finance. In this work, we discuss some upcoming active new research areas in application of quantum machine learning (QML) in finance. We discuss certain QML models that has become areas of active interest in the financial world for various applications. We use real world financial dataset and compare models such as qGAN (quantum generative adversarial networks) and QCBM (quantum circuit Born machine) among others, using simulated environments. For the qGAN, we define quantum circuits for discriminators and generators and show promises of future quantum advantage via QML in finance.

*Keywords—Quantum Generative Adversarial Network (qGAN), Quantum Circuits Born Machine (QCBM), QML*

## I. Quantum Computing in Finance

Quantum finance is an upcoming cross-discipline of financial engineering with the integration of quantum field theory, classical finance theory, computer science, and artificial intelligence (AI) technology [1]. Quantum machine learning (QML) is a cross-disciplinary subject made up of two of the most exciting research areas: quantum computing and classical machine learning. With recent advances of quantum computing technologies and the promised quantum advantage in QML, researchers have begun to consider how to utilize them in industries and a major focus area is aspects of finance (see review at []).

Since financial institutions are performing enormous tasks of numerical calculation in their daily works, promise of speed-up of such tasks by quantum computers are too enticing to ignore. For example, one of such tasks is pricing of financial derivatives: large banks typically have a huge number of derivatives written on various types of assets such as stock price, foreign exchange rate, interest rate, commodity and so on. Therefore, pricing of derivatives is an important issue for them. Another very important aspect of quantum finance is option pricing, closely related to financial derivatives [2]. Efforts of marrying quantum machine learning to stock market dynamics by leveraging the probabilistic nature of quantum computing and algorithms have been ongoing for forecasting and risk-analysis as well [3].

Quantum computers have been found to be useful in solving several well-known financial engineering problems such as Credit Risk Analysis (CRA) [4] . Simulation of portfolio loss distribution is achieved in classical computing by using the classical Monte Carlo method. Monte Carlo works by correlating underlying economic factors and associated asset relationships as a way to simulate all possible loss events. This is done by performing iterations over N samples taken from normal distribution of the data prior to averaging the several associated trajectories. Both Value at Risk (VaR) and Conditional Value at Risk (CVaR), critical properties of CRA evaluation, require large number of samples since both correspond to loss events that occur less frequently. This process tends to make classical Monte Carlo an expensive process, both computationally and time-wise. On the other hand, Quantum computers are deemed to be natural fit to solve Monte Carlo simulations - which are fundamental for derivatives pricing and related tasks, since they can be viewed as true random number generators. Quantum annealers have been found to be efficient in solving multi-period integer portfolio optimization problem, which is NP-Complete.

Quantum generative models, due to their inherent nature, promises to deliver quantum advantage on NISQ devices. This paper focuses on implementation of two quantum generative models as applied to financial problems: a) application of Quantum Generative Adversarial Networks (qGAN), a new and active research area that promises to be one where NISQ-based algorithms will be particularly fruitful for certain financial applications and b) Quantum Circuit Born Machine (QCBM). The motivation for this work is [5] which shows that Restricted Boltzmann Machines (RBM) outperforms parametric models for this type of financial dataset. This paper briefly describes classical and quantum option pricing, fundamentals of classical Generative Adversarial Networks (GAN) and qGAN, associated works done and implementation of qGAN for options pricing using real world financial dataset. Finally, we move onto the description and science of QCBM and implementation of it using real world dataset and discuss the results and future works for both schemes.

## II. GAN and qGAN

### A. Generative Adversarial Networks (GAN)

Generative models are machine learning (ML) models that employs very powerful statistical techniques. The advantage of GANs is that they can be trained in an unsupervised way. GANs were introduced in 2014 in a well cited paper [6] by Goodfellow et al. and tested initially on image datasets [7, 8], medicine [9, 10], Quantitative Finance [11], for portfolio optimization [12], fraud detection [13], trading model optimization [14] and generation of time series [15, 16].

GAN is an ML model used to train to generate data closely resembling the patterns and properties of a given dataset. This task is achieved by a GAN model by utilizing two neural networks (NN) as main components: a generator, which is a "generative" NN and a discriminator NN. The aim of the two NNs is to compete against each other. GAN implementations, however, have been found to often exhibit instability issues, vanishing gradient and mode collapse [17]. As per current literature, Wasserstein GANs, assuming the Wasserstein distance from optimal transport instead of the classical Jensen–Shannon Divergence, are still subject to slow convergence issues and potential instability [18]. There have been efforts to improve efficiency of GAN, for example, Lloyd and Weedbrook (2018) [19] and Dallaire-Demers and Killoran [20] introduced a quantum component to GANs, where the data consists of quantum states or classical data while the two players are equipped with quantum information processors.

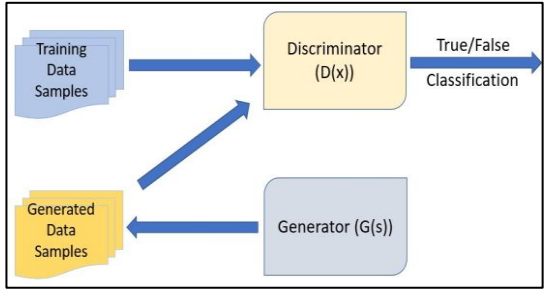

Fig 1. A classical GAN process. The Generator tries to create indistinguishable data samples from the training data. The differentiator tries to discern the generated data from the original data and outputs a True/False binary-classification.

A classical neural network, shown in Fig 2, has the following form:
$$f(\mathbf{x}, \mathbf{w_1}, \mathbf{w_2}, \dots) = \phi_2(\mathbf{w_2}\phi_1(\mathbf{w_1}\mathbf{x})),$$
where, $\phi_1, \phi_2$ etc. are nonlinear activation functions and $w_i$ are parameters defined as weight matrices. The model function $f$ is formed of concatenated linear models and functions. The concatenation can be repeated several folds [21].

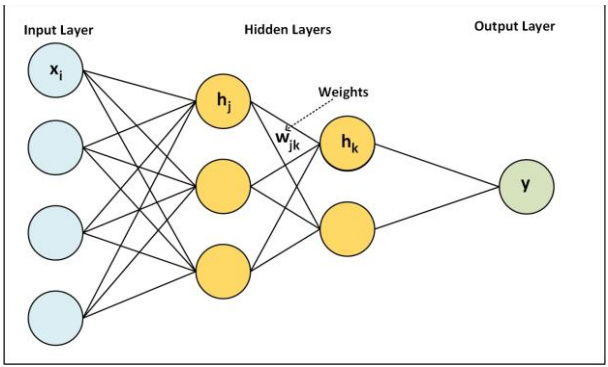

Fig 2. A classical neural network. $h_i$ are the hidden layers. $x_i$ and $y_i$ are the input and output layers.

For a given training data set of size $N, (x_i, y_i)_{i=1\dots N}$, the neural network builds a mapping between $(x_i)_{i=1\dots N}$ and $(y_i)_{i=1\dots N}$.

The GANs are trained via the following process: The generator and discriminator are initialized to some random configuration. The discriminator is trained to discern the real data from the output of the generator. The discriminator, trained in the previous step, tries to classify as many of the generated outputs as real. Then the discriminator is then trained on the newly generated dataset and the generator then re-trained to "fool" the discriminator. This process is iterated and ideally, at the each iteration it becomes harder for the discriminator to differentiate between the real data and the synthetic one from the generator. Eventually they reach an equilibrium where the generated data is discernible from the original data. If a classical training dataset $X = \{x_1 \dots x_{s-1}\}$ is sampled from an unknown probability distribution, $S$ is a set of seeds fed to the generator and $G_\theta$ and $D_\varphi$ are the actions of the generator and the discriminator with $\theta$ and $\varphi$ as parameters, then the discriminator training loss is given by

$$L_D = -\frac{1}{|X| + |S|} \left( \sum_{x \in X} \log D_\varphi(x) + \sum_{s \in S} \log(1 - D_\varphi(G_\theta(s))) \right) \quad (1)$$

In equation (1), $|X|$ and $|S|$ denote the respective size of sets. The job of the discriminator is to minimize this loss. The goal of training the generator is to match or "fool" the discriminator into classifying the generated data as real data, hence the goal of training the generator is to maximise $L_D$, which would be similar to minimizing $-L_D$. In equation (1), contribution of the first term is constant in the training of the generator as it does not depend on $G_\theta(s)$. Therefore, the goal of the generator training session is theoretically given by the minimization of the loss

$$L_G^* = \frac{1}{|S|} \sum_{s \in S} \log(1 - D_\varphi(G_\theta(s))) \quad (2)$$

However, practical considerations [6] prove that the more stable solution for $L_G$ is minimization of the loss function

$$L_G = -\frac{1}{|S|} \sum_{s \in S} \log(D_\varphi(G_\theta(s))) \quad (3)$$

The generator loss function $L_G$ decrease while the generator is trained making it more likely for the generated data to be erroneously classified as real data by the discriminator. It has been shown [6] that the discriminator assigns a value of $\sim 1/2$ to both $G_\theta(s)$ and $D_\varphi(x)$ at its optimal equilibrium since it is unable to distinguish between the real and generated data, giving $L_G = L_D = -\log 1/2 \approx 0.6931$.

*B. Quantum GAN (qGAN)*

A qGAN is a partial classical GAN with part of the model implemented via quantum neural networks (QNN). The training process of a qGAN is analogous to that of a classical GAN. qGANs are relatively very new and active research area, especially in financial domain. Research in qGANs are being done in several areas in finance, such as, to generate probability distributions for univariate distributions [22, 23] and for multivariate variate distributions [24, 25 QIF].

qGAN principles were introduced independently by Lloyd and Weedbrook [19] and also by Dallaire-Demers and Killoran [20]. The classical GAN problem was transformed by introducing density matrices for the qGAN version [19]. If some data is represented by a density matrix $\rho$ – which does not have to be a pure state – and a generator $G_q$ generating an output density matrix $\mu$, the discriminator tries to discern true data from the generated fake one. To do this, the discriminator makes a positive-operator-valued (POV) measurement with outcomes of True ($T$) or False ($F$). This gives the probability that measurement of a positive outcome given the true data is:

$$P(T|\rho) = \mathrm{Tr}(T\rho)$$

The probability that the measurement yields a positive answer is given by:

$$P(T|G) = \mathrm{Tr}(T\rho)$$

In this paper we explore the implementation of a qGAN for the uncertainty model on option pricing using real world data from Binance. The approach was first suggested by Zoufal et al. [26]. To address the qGAN scenario, a parametrised quantum generator is trained to work on a given $n$-qubit input state $|\psi\rangle$ to an $n$-qubit output state

$$G_{q\theta}|\psi\rangle_n = \sum_{i=0}^{2^n-1} \sqrt{p_\theta^i}|i\rangle_n \quad (4)$$

where $p_\theta^i$ is the output occurrence sampling probabilities of the basis state $i$. Starting from $n$-qubits, a quantum generator $G_{q\theta}$ takes the form of a multi-layer quantum neural network. For each layer $l \in \{1 \ldots L\}$, a unitary gate $U(\theta)_l$ acts on the $n$-qubits at the same time with dependence on hyperparameter $(\theta)_l$. In typical NISQ devices, entanglement is used for pairwise controlled gates to avoid high-order qubit gates in low circuit depths enforced by limited number of available qubits. Hence, for each layer $l$, $U$ is composed of a single or two-qubit gates only. The quantum generator is implemented by a variational form. The general form of each layer gate of the $L$-layer neural network is given by

$$U(\theta)_l = \left\{\bigotimes_{j=1}^{n} R_X(\theta_j) Q^{(j \bmod n)+1}(\theta_{j,imp})\right\}$$

$$\left\{\left(\bigotimes_{j=1}^{n} R_Z(\theta_{j,Z,l})\right)\left(\bigotimes_{j=1}^{n} R_X(\theta_{j,X,l})\right)\left(\bigotimes_{j=1}^{n} R_Z(\theta_{j,Y,l})\right)\right\}$$

where, $Q^j$ indicates that qubit $j$ is the control qubit and the gate action is on qubit $(j+1)$ and $((j \bmod n) + 1) = 1 + j$ when $j \in \{1 \ldots, n-1\}$ and is 1 when $j = n$. The total number of hyperparameters are given by $5nl$, with $5n$ per layer. The discriminator itself, generally may or may not be quantum in nature depending on the nature of the problem. In our particular case, the discriminator is a classical binary classifier.

### III. METHODS AND RESULTS

This section addresses the implementation methods and discussions about the results.

### A. qGAN

The qGAN is made up of a quantum circuit that will be able to prepare a 1-qubit state $|\psi_1\rangle$, also called the "true" state, that we want the qGAN to learn. This circuit runs on the first qubit of the model. $|\psi_1\rangle$ will act as the quantum training data for the training circuit. The circuit will try to produce as many $|\psi_1\rangle$ states as possible for the training process. The quantum generator, implemented via a variational form with trainable parameters, will also run on the first qubit of the device in order to prepare a state similar to $|\psi_1\rangle$ on the first qubit. The quantum discriminator will try to distinguish between the two states. In this case the discriminator used is a classical binary classifier. The goal is for the quantum generator to learn a representation for the probability distribution underlying the training data. In this work we explored the creation of a qGAN for the uncertainty model on option pricing.

Data: Real life data were used via Binance API call for five different crypto currencies with more than 5000 samples: BNBBTC, ETHBTC, LTCBTC, NEOBTC, QTUMETH (see Fig A2 for sample data). Data samples smaller than 5% percentile and bigger than 95% percentile were discarded to reduce the number of qubits required for a reasonable representation of the distribution. The optimization scheme used data batches of 1,000 and was run for 2000 training epochs mainly utilizing IBM's Qiskit 0.39 and PyTorch .

First the training data $X$ was loaded. The trained quantum generator can, be used to load a quantum state which is an approximate model of the target distribution and its goal is to correspond to the $n$-qubit quantum state $G_{q\theta}|\psi\rangle_n = \sum_{i=0}^{2^n-1} \sqrt{p_\theta^i}|i\rangle_n$. This required a mapping of the samples from the multivariate normal distribution to discrete values. The number of values that can be represented depends on the number of qubits used for the mapping, i.e, the data resolution is defined by the number of qubits. Hence, for 5 assets and 3 qubits to represent one feature, there were $2^5 = 32$ discrete values. Then Then, the discretization of the training data was verified by passing data resolution as an array [3, 3, 3, 3, 3] which defined the number of qubits used to represent each data dimension. As such, $2^5 = 32$ discrete values and all training samples should fall under one of these discrete values. Then, histograms in Fig. A1 of the appendix were generated for each random variable the number of bins were increased and to make the actual discretization visible.

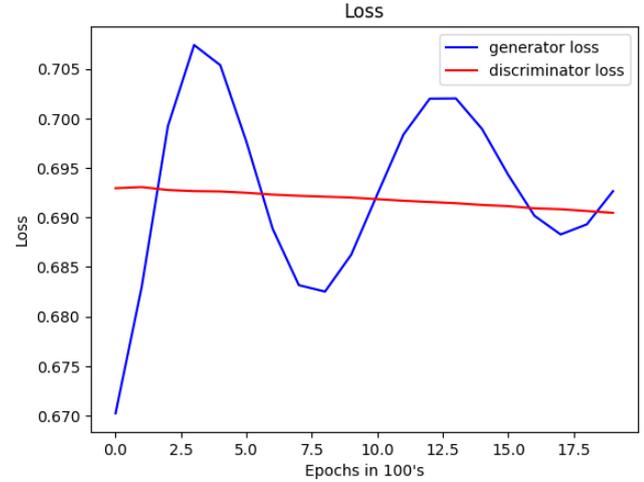

Fig 4. Quantum generator vs. discriminator loss function

Modeling, conversion of data arrays into tensors and creation of a data loader from the training data was achieved by using PyTorch 2.0. Then a quantum instance was craeted for the training where the batch size defines the number of shots using Qiskit Aer. For the qGAN implementation, a parametrized quantum circuit $G_{q\theta}$ to create the quantum generator. The discriminator was created with a classical Neural Network approach with its output designed to ideally identify fake from real data as a binary classification.

An Adam optimizer was used for the gradient based optimization process. Fig 3 below shows the generator circuit.

The loss functions of the generator and discriminator compete against each other reaching the nash-equilibrium. In our case, the generator loss functions are parametrized by $\theta$ those of the discriminator are defined by $\varphi$. The discriminator, which has to differentiate and classify two data points $|\psi_1\rangle$, labeled '1' and the generated state $|\psi_g\rangle$ with a label '0', has a loss function for the quantum format given by,

$$L_{q\varphi} = -0.5[\log(1 - D|\psi_g\rangle) + \log(D|\psi_1\rangle)]$$

and the generator loss function is given by

$$L_{G\theta} = -\log(D|\psi_g\rangle)$$

The generator vs. discriminator loss function for the option pricing uncertainty of the Binance data is shown in Fig. 4 below.

The generator not only takes into consideration the singular evolution of an option, but also the correlated evolution of other market assets.

As expected, the discriminator assigns a value of $\sim 1/2$ to both $G_\theta(s)$ and $D_\varphi(x)$ at its *optimal equilibrium* where the two loss functions intersect since it is unable to distinguish between the real and generated data, giving $L_G = L_D = -\log 1/2 \approx 0.6931$. Lastly, the fidelity relationship between the data and the data types of the generator and discriminator were investigated as shown in Fig. 5 and as expected showed

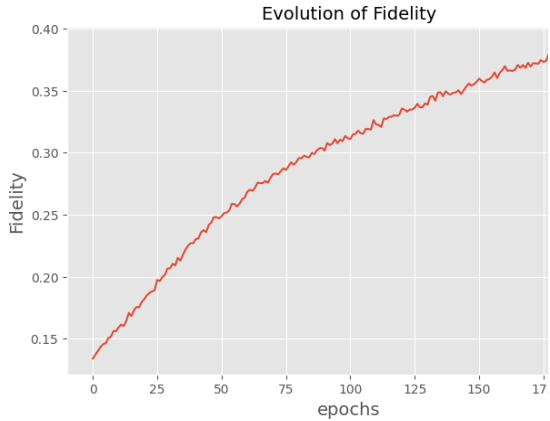

Fig 5. Evolution of fidelity.

a steadily increasing rise in value with an increase in iterations.

### B. QCBM

A main focus of machine learning is probabilistic modeling. A probability distribution is obtained from a finite set of samples. If the training process is successful, the learned distribution P(x) has sufficient similarity to the actual distribution of the data that it can make correct predictions about unknown situations.

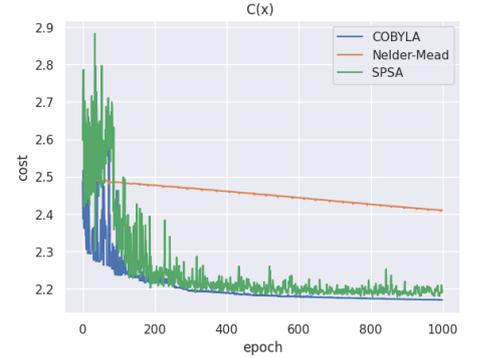

Fig 8. Comparison of optimizers. Cobyla gave the best results.

In order to build a QCBM, we need to specify the number of layers, type of fixed gates for each layer and type of adjustable gates. Initially a histogram (Fig. 6) was generated to prove that the state vector satisfies the characteristics of equation (5).

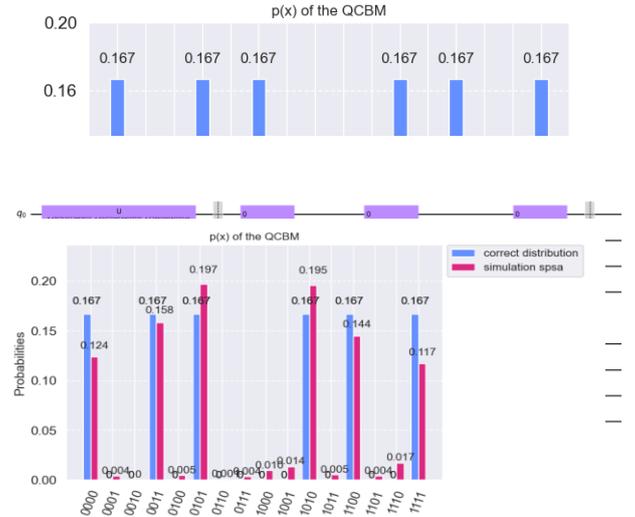

Fig 9. Comparison of real and simulated distribution with SPSA.

Fig 7 shows the QCBM circuit decomposition as generated by Qiskit. Circuit decomposition of the QCBM was done to verify that the same circuit is repeated in layer 1 and layer 3, using the odd layer, and the second circuit is the even layer. The QCBM circuit was constructed as shown in Fig. 7.

For the dataset, one of the goal is to obtain an approximation to the target probability distribution px_output or $P(x)$. This was achieved with a quantum circuit parametrized by $\theta$. The layer were defined as 5. Minimization of the cost function Minimization of cost function $\mathbb{C}(\boldsymbol{\theta}) = \frac{1}{-d}\sum_{d=1}^{5}\ln(P(\mathbf{x}^d))$, where 'd' is the number of layers (5 in this case) and $P(\mathbf{x}) = |\langle \mathbf{x}|\psi(\boldsymbol{\theta})\rangle|^2$ is the probability. A simple variant of $\mathbb{C}(\boldsymbol{\theta})$ was used $\mathbb{C}(\boldsymbol{\theta}) = -\frac{1}{d}\sum_{d=1}^{5}\ln(\max(\epsilon, P_\theta(\mathbf{x}^d))$, where $\epsilon > 0$ is a small parametric perturbation. A method called born_machine(params) was defined for this purpose. A comparison of e optimzers were done between Cobyla, SPSA and Nelder-Mead as shown in Fig. 8. Cobyla was found to be marginally better than SPSA.

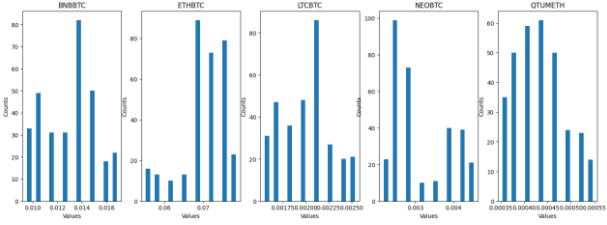

Fig A1. Discrete values and training samples for qGAN

| Asset | BNBBTC | ETHBTC | LTCBTC | NEOBTC | QTUMETH |
|---|---|---|---|---|---|
| Closing time | | | | | |
| 1647215999999 | 0.009564 | 0.066587 | 0.002690 | 0.000503 | 0.002209 |
| 1647302399999 | 0.009411 | 0.065276 | 0.002661 | 0.000493 | 0.002220 |
| 1647388799999 | 0.009460 | 0.066640 | 0.002722 | 0.000495 | 0.002188 |
| 1647475199999 | 0.009374 | 0.067468 | 0.002707 | 0.000502 | 0.002167 |
| 1647561599999 | 0.009574 | 0.068726 | 0.002694 | 0.000505 | 0.002107 |
| ... | ... | ... | ... | ... | ... |
| 1678406399999 | 0.013607 | 0.070597 | 0.003767 | 0.000486 | 0.001774 |
| 1678492799999 | 0.013766 | 0.070788 | 0.003544 | 0.000487 | 0.001759 |
| 1678579199999 | 0.013485 | 0.071945 | 0.003382 | 0.000466 | 0.001672 |
| 1678665599999 | 0.013099 | 0.071836 | 0.003466 | 0.000468 | 0.001724 |
| 1678751999999 | 0.012771 | 0.069387 | 0.003378 | 0.000458 | 0.001702 |

Fig A2. Sample real life data from Binance

Finally, a comparison of the correct probability distribution and simulation by SPSA was done and SPSA simulation appeared to be close to the real distribution as shown in Fig. 9. Comparison of the real data with Cobyla had wider gaps with the real fdata even though Cobyla, as an optimizer, gave better results.

## IV. CONCLUSION

A study of implementations of qGAN and QCBM were done using real life Binance data on Qiskit simulator from IBM. In the qGAN case, the loss functions of the discriminator and generator behaved a expected. The qGAN produced expected results in far less iterations than it would take a classical GAN. qGAN is a new and exciting area of research that promises robust advantage in treating and modelling financial data.

The QCBM implementation with the same data produced results with SPSA optimization which were closely matched with the real data. This shows that both qGANs and QCBM promises advantage from quantum machine learning perspective in financial data Science. In recent times promising quantum advantage have been reported by combining QCBM with annealing [34] . Further work is required in this exciting area.

## V. APPENDIX